\documentclass[12pt]{article}
\usepackage{amssymb}
\usepackage{amsmath,,calrsfs}
\usepackage{bbold}
\usepackage[dvips]{epsfig}
\allowdisplaybreaks

\newcommand{\be}{\begin{equation}}
\newcommand{\ee}{\end{equation}}
\def\bea{\begin{eqnarray}}
\def\eea{\end{eqnarray}}

\newcommand{\bn}{\begin{eqnarray}}
\newcommand{\en}{\end{eqnarray}}

\newcommand{\p}{\partial}

\newcommand{\nn}{\nonumber}

\newcommand{\no}{\noindent}

\newcommand{\s}{\,\,\,\,}
\def\bea{\begin{eqnarray}}
\def\eea{\end{eqnarray}}

\newcommand{\beq}{\begin{eqnarray}}
\newcommand{\eeq}{\end{eqnarray}}
\begin{document}

\title{\textbf{Unitarity of Singh-Hagen model in $D$ dimensions }}
\author{E. L. Mendon\c ca \footnote{eliasleite@feg.unesp.br}, R. Schimidt Bittencourt\footnote{raphael.schimidt@unesp.br} \\
\textit{{UNESP - Campus de Guaratinguet\'a - DFQ} }\\
\textit{{Av. Dr. Ariberto Pereira da Cunha, 333} }\\
\textit{{CEP 12516-410 - Guaratinguet\'a - SP - Brazil.} }\\
}
\date{\today}
\maketitle

\begin{abstract}
The particle content of the Singh-Hagen model ($SH$) in $D$ dimensions is revisited. We suggest a complete set of spin-projection operators acting on totally symmetric rank-3 fields. We give a general expression for the propagator and determine the coefficients of the $SH$ model confirming previous results of the literature. Adding source terms we provide an unitarity analysis in $D$ dimensions. In addition we have also analyzed the positivity of the massless Hamiltonian.
\end{abstract}
\newpage

\section{Introduction}
The suggestion of a free theory describing higher spin particles dates back to 1936 by Dirac \cite{dirac} and 1939 by Fierz and Pauli $(FP)$ \cite{FP}. As fundamental assumptions, such theories should be invariant under Poincaré transformations and at the same time guarantee the energy positivity. Particularly in the case of higher spin theories, positivity deserves special attention
since attempting to describe such particles, we are faced with extra propagation modes with spins lower than those we would like to describe, such modes may be ghosts in some cases. In order to remove the spurious degrees of freedom, one needs the addition of auxiliary fields, which sometimes makes the analysis of the equations of motion truly complicated.

From the experience with lower spins one knows that the particle content of some theory may be directly obtained by calculating the propagator, but in order to obtain it, one needs to construct a complete basis of spin-projection operators. Strictly speaking about bosonic examples, it is quite simple to obtain the propagator of a rank-one field theory with the help of the transverse $\theta_{\mu\nu}=\eta_{\mu\nu}-\omega_{\mu\nu}$ and longitudinal $\omega_{\mu\nu}=\p_{\mu}\p_{\nu}/\Box$ operators. By mean of these projectors Barnes and Rivers have introduced \cite{Rivers} a complete set of spin-projection operators which allows us to determine the particle content of a given rank-two field theory (a slightly different basis is also used by \cite{niew}). Some extensions of this set of projectors are given at \cite{Pires, Helay} where a new class of projection operators for three-dimensional models are constructed.

The spin-3 case is the simplest bosonic example of a higher spin theory. A model of second order in derivatives which describe a massive spin $3$ particle is given by Singh and Hagen \cite{SH}. Here we revisit the particle content of this model in $D$ dimensions by suggesting a complete set of spin-projection operators, our results in some sense generalizes the discussion cared out by \cite{Reisen} and are in agreement with those results obtained by \cite{deseryang}. We also provide a unitarity analysis of such a model by adding source terms in order to verify the sign of the imaginary part of the residue of the transition amplitude saturated in the sources.
For the massless case of the $SH$ theory we have obtained the canonical Hamiltonian as well as the constraints of the theory in $D$ dimensions. By checking that they are first class constraints we demonstrate that the system describes the correct number of degrees of freedom. Finally by using the constraints as strong equalities we provide the reduced Hamiltonian in terms of spin-projection operators demonstrating that the model carries only spin-3 particles and that it is positive definite.  

\section{Rank-3 spin-projection operators}

In the $SH$ model the spin-3 field is a totally symmetric field $h_{\mu\nu\lambda}$ with the trace given by $h_{\lambda}=\eta^{\mu\nu}h_{\mu\nu\lambda}$. Along of this work we have used the mostly plus metric $(-,+,+,..)$. At least in $D=3+1$ we should expect six projection operators once the field  $h_{\mu\nu\lambda}$ belongs to the representation of the Lorentz group given by $ (1/2;1/2)\otimes(1/2;1/2)\otimes(1/2;1/2)=(3/2;3/2)\oplus(1/2;1/2)$. They corresponds to the unique spin-3 sector given by the symmetric, transverse and traceless part of $h_{\mu\nu\lambda}$, one spin-2 sector given by the divergence $\p^{\mu}h_{\mu\nu\lambda}$, two spin-1 sectors contained at the double divergence $\p^{\mu}\p^{\nu}h_{\mu\nu\lambda}$ and trace $\eta^{\mu\nu}h_{\mu\nu\lambda}$ and two spin-0 sectors given by the triple divergence $\p^{\mu}\p^{\nu}\p^{\lambda}h_{\mu\nu\lambda}$ and the divergence of the trace $\p^{\mu}h_{\mu}$. Such projectors in this specific dimension are given for example in \cite{Reisen} \footnote{A slightly different basis can also be found for example in \cite{Chiang} and a set of semi-projectors operators was explored by \cite{Nelipa}. Both basis are not convenient to our purposes, besides they are also in $D=3+1$. A recent development has also bee achieved for the rank-s case in $D$ dimensions at \cite{russo}}. Aiming the construction of them in $D$ dimensions we notice that the trace of $\theta_{\mu\nu}=\eta_{\mu\nu}-\omega_{\mu\nu}$ and $\omega_{\mu\nu}={\p_{\mu}\p_{\nu}}/{\Box}$ are respectively $D-1$ and 1. Then one can generalize that results to $D$ dimensions as follows:

\bea
(P^{(3)}_{11})^{\mu\nu\rho}_{\alpha\beta\gamma} &=&  \theta^{(\mu}_{(\alpha} \theta^\nu_\beta \theta^{\rho)}_{\gamma)} - (P^{(1)}_{11})^{\mu\nu\rho}_{\alpha\beta\gamma},\label{first} \\
(P^{(2)}_{11})^{\mu\nu\rho}_{\alpha\beta\gamma} &=&  3 \theta^{(\mu}_{(\alpha} \theta^\nu_\beta \omega^{\rho)}_{\gamma)} - (P^{(0)}_{11})^{\mu\nu\rho}_{\alpha\beta\gamma},  \\
(P^{(1)}_{11})^{\mu\nu\rho}_{\alpha\beta\gamma} &=&  \frac{3}{(D+1)}  \theta^{(\mu\nu} \theta_{(\alpha\beta}  \theta^{\rho)}_{\gamma)},\\
(P^{(1)}_{22})^{\mu\nu\rho}_{\alpha\beta\gamma} &=&  3 \theta^{(\mu}_{(\alpha} \omega^\nu_\beta \omega^{\rho)}_{\gamma)},  \\
(P^{(0)}_{11})^{\mu\nu\rho}_{\alpha\beta\gamma} &=&  \frac{3}{(D-1)}  \theta^{(\mu\nu} \theta_{(\alpha\beta}  \omega^{\rho)}_{\gamma)}, \label{p110} \\
(P^{(0)}_{22})^{\mu\nu\rho}_{\alpha\beta\gamma} &=&   \omega_{\alpha\beta}
\omega^{\mu\nu} \omega^{\rho}_{\gamma} \label{p220} \eea

\no Notice that, here the parenthesis means normalized symmetrization, taking for example:
\be \theta^{(\mu}_{(\alpha} \theta^\nu_\beta \theta^{\rho)}_{\gamma)}=\frac{1}{6}(
\theta^{\mu}_{\alpha} \theta^\nu_\beta \theta^{\rho}_{\gamma}+
\theta^{\rho}_{\alpha} \theta^\nu_\beta \theta^{\mu}_{\gamma}+
\theta^{\nu}_{\alpha} \theta^\mu_\beta \theta^{\rho}_{\gamma}+
\theta^{\rho}_{\alpha} \theta^\mu_\beta \theta^{\nu}_{\gamma}+
\theta^{\nu}_{\alpha} \theta^\rho_\beta \theta^{\mu}_{\gamma}+
\theta^{\mu}_{\alpha} \theta^\rho_\beta \theta^{\nu}_{\gamma}).\ee

As a requirement the basis must be orthonormal and the projectors idempotents:

\be
P_{ij}^{(s)}P_{kl}^{(r)}=\delta^{sr}\delta_{jk}P_{il}^{(s)}.\label{algebra}
\ee

\no  In our notation, the superscripts $(r)$ and $(s)$ denote the spin subspace, while subscripts $i,j,k,l$ works to distinguish between projectors and transition operators. Once for example: $i=j$ or $k=l$ we have a projector, while if $i\neq j$ or $k\neq l$ we have a transition operator. In addition, the subscripts work in order to count the number of projectors of a given spin subspace, for example in the subspace of spin $0$ we have two projectors represented by the combinations $i=1, j=1$ and $i=2, j=2$. The set of projectors obey the following mathematical identity:
\be
\sum_{i,s} P^{(s)}_{ii}  = \mathbb{1}, \label{id}
\ee

 \no where $\mathbb{1}$ stands for the  symmetric rank-3 identity operator, i.e.:
\bea \mathbb{1}^{\mu\nu\rho}_{\alpha\beta\gamma} = 
\delta^{(\mu}_{(\alpha} \delta^\nu_\beta \delta^{\rho)}_{\gamma)}.\eea

Finally, the transition operators $P^{(s)}_{{ij}}$ are given by:
\bea
(P^{(1)}_{{12}})^{\mu\nu\rho}_{\alpha\beta\gamma} &=&  \frac{3}{\sqrt{(D+1)}}   \theta_{(\alpha\beta} \theta^{(\rho}_{\gamma)} \omega^{\mu\nu)},  \\
(P^{(1)}_{{21}})^{\mu\nu\rho}_{\alpha\beta\gamma} &=&  \frac{3}{\sqrt{(D+1)}}   \theta^{(\mu\nu} \theta^{\rho)}_{(\gamma} \omega_{\alpha\beta)},  \\
(P^{(0)}_{{12}})^{\mu\nu\rho}_{\alpha\beta\gamma} &=&  \frac{3}{\sqrt{3(D-1)}}   \theta_{(\alpha\beta} \omega^{(\mu\nu} \omega^{\rho)}_{\gamma)},   \\
(P^{(0)}_{{21}})^{\mu\nu\rho}_{\alpha\beta\gamma} &=& 
\frac{3}{\sqrt{3(D-1)}}   \theta^{(\mu\nu} \omega_{(\alpha\beta} \omega^{\rho)}_{\gamma)}. \eea

We have to say that the transition operators also satisfy the algebra given by (\ref{algebra}). They are not necessary to complete the identity in (\ref{id}), however we do need them in order to expand the sandwiched operator between two rank-3 fields in a bilinear lagrangian.
 
\section{On the coefficients of the spin-3 Singh-Hagen theory}
As showed in \cite{kobayashi1, kobayashi2} the description of a massive spin-3 particle in terms of a totally symmetric field with a second order lagrangian, requires the introduction of an auxiliary field. The simplest way to introduce it is by associating the totally symmetric field with a scalar field. In \cite{hds3} one of us in collaboration has observed that even for higher derivative descriptions of dublets of spin-3 it is necessary the addition of such auxiliaries. Recently, however, the authors in \cite{ddhd} have verified that they are not necessary for a higher derivative self-dual description of a massive spin-3 singlet in $D=2+1$ dimensions.

In the following lines we use the operators to revisit, as an example, the particle content of the $SH$ model coupled in the simplest way to a scalar auxiliary field $W$. Let us suppose all the coefficients are undetermined  and given by $a,b,...t$, then we have:

\begin{align}
\mathcal{L} = & a\; h_{\mu\nu\rho} \Box h^{\mu\nu\rho} + b\; h_{\mu\nu\rho} \p^\mu \p_\alpha h^{\alpha\nu\rho} + c\; h_{\nu} \p^\nu \p_\mu h^{\mu} + d\; h_{\nu} \Box h^{\nu} + e\; h_{\rho} \p_\mu \p_\nu h^{\mu\nu\rho} \nn \\ 
& + fm^2\; h_{\mu\nu\rho} h^{\mu\nu\rho} + gm^2\; h_{\mu} h^{\mu} + i\; W \Box W + jm^2\;W^2 + tm\; h_{\mu}\p^\mu W, \label{14}
\end{align}

In the appendix we give explicit expressions of the bilinear form of each term. By collecting all of them the lagrangian density can be written as:
\bea
\mathcal{L} = h_{\mu\nu\rho} \Theta^{\mu\nu\rho}_{\kappa\lambda\sigma} h^{\kappa\lambda\sigma} + W \varPhi W + h_{\mu\nu\rho} T^{\mu\nu\rho}W,
\label{lagop}\eea
where
\bea
\Theta^{\mu\nu\rho}_{\kappa\lambda\sigma}&=& (a\Box + f m^2) \left( P^{(3)}_{11}\right)^{\mu\nu\rho}_{\kappa\lambda\sigma}  + \left[ \left( a + \frac{b}{3} \right)\Box  + f m^2\right] \left( P^{(2)}_{11}\right)^{\mu\nu\rho}_{\kappa\lambda\sigma} \nn \\
&+& \left\lbrace \left[ a + \frac{(D+1)}{3}d \right]\Box  + \left[ f + \frac{(D+1)}{3}g \right]m^2 \right\rbrace  \left( P^{(1)}_{11}\right)^{\mu\nu\rho}_{\kappa\lambda\sigma} \nn \\
&+& \left[ \left( a + \frac{2b+d+e}{3} \right)\Box  + \left( f + \frac{g}{3} \right)m^2 \right]  \left( P^{(1)}_{22}\right)^{\mu\nu\rho}_{\kappa\lambda\sigma} \nn \\
&+& \left\lbrace \left[ a + \frac{b+(D-1)(c+d)}{3} \right]\Box  + \left[ f + \frac{(D-1)}{3}g \right]m^2 \right\rbrace  \left( P^{(0)}_{11}\right)^{\mu\nu\rho}_{\kappa\lambda\sigma} \nn \\
&+& \left[ \left( a + b + c + d + e \right)\Box  + \left( f + g \right)m^2 \right]  \left( P^{(0)}_{22}\right)^{\mu\nu\rho}_{\kappa\lambda\sigma} \nn \\
&+&\frac{\sqrt{D+1}}{3} \left[  \left(d+\frac{e}{2}\right) \Box  +  g m^2 \right] \left( P^{(1)}_{12} + P^{(1)}_{21} \right)^{\mu\nu\rho}_{\kappa\lambda\sigma} \nn \\
&+& \frac{\sqrt{3(D-1)}}{3}\left[  \left(c+d+\frac{e}{2}\right) \Box  +  gm^2 \right]  \left( P^{(0)}_{12} + P^{(0)}_{21} \right)^{\mu\nu\rho}_{\kappa\lambda\sigma}.
\eea
\bea
\varPhi = (i\Box + j m^2),
\eea
\bea
T^{\mu\nu\rho} = \frac{t m}{3}(\eta^{\mu\nu}\p^\rho + \eta^{\nu\rho}\p^\mu + \eta^{\rho\mu}\p^\nu).
\eea
\no Integrating over the scalar field $W$ we then have the non-local lagrangian given by:

\bea
\mathcal{L}=h_{\mu\nu\rho} G^{\mu\nu\rho}_{\alpha\beta\gamma} h^{\alpha\beta\gamma},
\eea

\no where the operator $G$ is written as:

\bea
G^{\mu\nu\rho}_{\kappa\lambda\sigma}\equiv  \Theta^{\mu\nu\rho}_{\kappa\lambda\sigma} + \frac{1}{4\varPhi} T^{\mu\nu\rho} T_{\kappa\lambda\sigma}.
\eea

In order to determine the coefficients $a,b,...t$ we now take the equations of motion with respect to the symmetric field $h^{\mu\nu\lambda}$ and then to select the spin-3, spin-2, spin-1 and spin-0 sectors we apply the spin-projection operators on these equations, which give us:
\begin{align}
\left( P_{11}^{(3)} \right)^{\alpha\beta\gamma}_{\mu\nu\rho} \left( G h \right)^{\mu\nu\rho} = & \left( a\Box + f m^2\right) \left( P_{11}^{(3)} h \right)^{\alpha\beta\gamma}, \label{E1} \\ \nn \\
\left( P_{11}^{(2)} \right)^{\alpha\beta\gamma}_{\mu\nu\rho} \left( G h \right)^{\mu\nu\rho} = & \left[ \left( a + \frac{b}{3} \right) \Box + f m^2\right] \left( P_{11}^{(2)} h \right)^{\alpha\beta\gamma}, \label{E2} \\ \nn \\
\left( P_{11}^{(1)} \right)^{\alpha\beta\gamma}_{\mu\nu\rho} \left( G h \right)^{\mu\nu\rho} = &   \left\lbrace\left[ a + \frac{(D+1)}{3} d \right] \Box + \left[ f + \frac{(D+1)}{3} g \right]m^2 \right\rbrace \left( P_{11}^{(1)} h \right)^{\alpha\beta\gamma} \nn \\ \nn \\
+ & \frac{\sqrt{D+1}}{3} \left[ \left(d + \frac{e}{2} \right) \Box + g m^2 \right] \left( P_{12}^{(1)} P_{22}^{(1)} h \right)^{\alpha\beta\gamma}, \label{E3} \\ \nn \\
\left( P_{22}^{(1)} \right)^{\alpha\beta\gamma}_{\mu\nu\rho} \left( G h \right)^{\mu\nu\rho} =& \frac{\sqrt{D+1}}{3} \left[  \left(d + \frac{e}{2} \right) \Box +   g  m^2 \right] \left( P_{21}^{(1)} P_{11}^{(1)} h \right)^{\alpha\beta\gamma} \nn \\ \nn \\
+& \left[ \left( a + \frac{2b+d+e}{3} \right) \Box + \left(f + \frac{g}{3} \right)m^2 \right] \left( P_{22}^{(1)} h \right)^{\alpha\beta\gamma}, \label{E4} \\ \nn \\
\left( P_{11}^{(0)} \right)^{\alpha\beta\gamma}_{\mu\nu\rho} \left( G h \right)^{\mu\nu\rho} = & \left\lbrace \left[ a + \frac{b}{3} + \frac{(D-1)(c+d)}{3} \right] \Box + \left[ f + \frac{(D-1)}{3} g \right]m^2 \right\rbrace \left( P_{11}^{(0)} h \right)^{\alpha\beta\gamma} \nn \\ \nn \\
+& \frac{\sqrt{3(D-1)}}{3} \left[ \left(c + d + \frac{e}{2} \right) \Box +   g m^2 \right] \left( P_{12}^{(0)} P_{22}^{(0)} h \right)^{\alpha\beta\gamma} \nn \\ \nn \\
+  \frac{t^2m^2\Box}{12(i\Box+jm^2)} &  \left[(D-1)\left( P_{11}^{(0)} h \right)^{\alpha\beta\gamma}  + \sqrt{3(D-1)}\left( P_{12}^{(0)} P_{22}^{(0)} h \right)^{\alpha\beta\gamma} \right], \label{E5} \\ \nn \\
\left( P_{22}^{(0)} \right)^{\alpha\beta\gamma}_{\mu\nu\rho} \left( G h \right)^{\mu\nu\rho} = &\frac{\sqrt{3(D-1)}}{3} \left[ \left(c + d + \frac{e}{2} \right) \Box +   g m^2 \right] \left( P_{21}^{(0)} P_{11}^{(0)} h \right)^{\alpha\beta\gamma}  \nn \\ \nn \\
+&  \left[ \left( a + b + c + d + e \right) \Box + \left(f + g \right)m^2 \right] \left( P_{22}^{(0)} h \right)^{\alpha\beta\gamma} \nn \\ \nn \\
+  \frac{t^2m^2\Box}{12(i\Box+jm^2)} &  \left[\sqrt{3(D-1)}\left( P_{21}^{(0)} P_{11}^{(0)} h \right)^{\alpha\beta\gamma}  + 3\left( P_{22}^{(0)} h \right)^{\alpha\beta\gamma} \right]. \label{E6}
\end{align} \\

 \no Once we want to have only the spin-3 propagation we need to handle the equations from (\ref{E1}) to (\ref{E6}) in order to have a Klein-Gordon equation in the spin-3 sector and to kill all the subsidiary conditions propagating lower spins, which can be done by setting to zero the coefficients multiplying the d`Alembertian in such sectors. After manipulating with the system of equations given above one can then find the coefficients and their relations, which  after substituting back in the lagrangian density give us:
\begin{align}
\mathcal{L} = & a\; h_{\mu\nu\rho} \Box h^{\mu\nu\rho} - 3a\; h_{\mu\nu\rho} \p^\mu \p_\alpha h^{\alpha\nu\rho} - \frac{3a}{2}\; h_{\nu} \p^\nu \p_\mu h^{\mu} - 3a\; h_{\nu} \Box h^{\nu} + 6a\; h_{\rho} \p_\mu \p_\nu h^{\mu\nu\rho} \nn \\ 
& - am^2\; h_{\mu\nu\rho} h^{\mu\nu\rho} + 3am^2\; h_{\mu} h^{\mu} - \frac{1}{3a}\frac{(D-1)}{(D-2)} \; t^2 \; W \Box W + \frac{1}{2a}\left(\frac{D}{D-2}\right)^2 t^2 m^2\;W^2 \nn \\
& + tm\; h_{\mu}\p^\mu W. \label{LSH}
\end{align}

We end up with two undetermined coefficients $a$ and $t$, however, notice that if we redefine $h\to h/\sqrt{2a}$ and $W\to \sqrt{2a}W/t$ they are completely eliminated of the action. Besides,  we notice that the dimensional dependence in the coefficients of $W\Box W$ and $W^2$ are exactly the same the authors have obtained in \cite{deseryang}. This is precisely the $SH$ model in $D$ dimensions and in the next section we are going to analyze its unitarity. The presence of auxiliary fields in higher spin theories is always the reason for difficulties when analyzing the equations of motion. We have noticed that if one takes the lagrangian given by (\ref{14}) and eliminate the scalar field ad hoc, we can then perform the analysis of the equations of motion by projecting the spin sectors and to conclude that we have a Klein-Gordon equation to the spin-3 mode if and only if $D=2$. In other words, in this specific dimension, we do not need the presence of auxiliary fields. This is an expected result once there is no reason to think about spins in such a dimension.

\section{Unitarity of the spin-3 Singh-Hagen model}

Here we start by supposing that one can integrate over the auxiliary scalars obtaining a non-local lagrangian which however can be put in a bilinear form. Then the sandwiched operator can be expanded in terms of the orthonormal basis introduced before. As a warm-up exercise, we could take a general bilinear lagrangian given by:
\bea
\mathcal{L}=h_{\mu\nu\rho} G^{\mu\nu\rho}_{\alpha\beta\gamma} h^{\alpha\beta\gamma},
\eea
where $G^{\mu\nu\rho}_{\alpha\beta\gamma}$ is an operator that can be written in terms of $P_{ij}^{(s)}$. From now on we have suppressed the indices for sake of simplicity in all the results.
\begin{align}
G=&A P_{11}^{(3)} + B P_{11}^{(2)} + C P_{11}^{(1)} + D P_{22}^{(1)} + E (P_{12}^{(1)} + P_{21}^{(1)}) \nonumber \\ & + F P_{11}^{(0)} + H P_{22}^{(0)} + I (P_{12}^{(0)} + P_{21}^{(0)}),
\end{align}
and suppose the inverse of $G$ given by:
\begin{align}
G^{-1}=&J P_{11}^{(3)} + K P_{11}^{(2)} + L P_{11}^{(1)} + M P_{22}^{(1)} + N (P_{12}^{(1)} + P_{21}^{(1)}) \nonumber \\ & + Q P_{11}^{(0)} + R P_{22}^{(0)} + S (P_{12}^{(0)} + P_{21}^{(0)}).
\end{align}

\no The coefficients $A, B,...S$ are completely arbitrary, but once we know that $GG^{-1}=\mathbb{1} $ we can relate them through the general results below:

\bea
J&=&\frac{1}{A}; \s\s\s K=\frac{1}{B}; \\
L&=&\frac{D}{DC-E^2}; \s\s\s M=\frac{C}{DC-E^2}; \s\s\s N=-\frac{E}{DC-E^2}; \\
Q&=&\frac{H}{HF-I^2}; \s\s\s R=\frac{F}{HF-I^2}; \s\s\s\s S=-\frac{I}{HF-I^2}.
\eea 

Such results are useful for any spin-3 model without parity breaking. This is precisely the case of the lagrangian density we have found at (\ref{LSH}) after field redefinitions.   One can verify that the operator $G$ in this case is given by:
\bea
G &=& \frac{(\Box -  m^2)}{2} \left( P^{(3)}_{11}\right)  -\frac{m^2}{2}  \left( P^{(2)}_{11}\right)- \frac{D}{2} \left(\Box  - m^2 \right)  \left( P^{(1)}_{11}\right) \nn \\
&-&\frac{1}{2} \left[ \frac{3(D-1)\Box}{2}   -  (D-2)m^2 \right]  \left( P^{(0)}_{11}\right) \nn \\
&-&\frac{1}{2} \left[\frac{\Box}{2}-2m^2\right]  \left( P^{(0)}_{22}\right)+  \frac{1}{2}{\sqrt{D+1}m^2}   \left( P^{(1)}_{12} + P^{(1)}_{21} \right) \nn \\
&-&  \frac{1}{2}{\sqrt{3(D-1)}}\left( \frac{\Box}{2}  -  m^2 \right)  \left( P^{(0)}_{12} +  P^{(0)}_{21} \right) \nn \\
&+& \frac{ m^2\Box}{12(i\Box+jm^2)}\left[ (D-1) \left( P^{(0)}_{11} \right) + 3\left( P^{(0)}_{22} \right)\right]+\frac{ m^2\sqrt{3(D-1)}\Box}{12(i\Box+jm^2)}  \left( P^{(0)}_{12} + P^{(0)}_{21} \right).\label{G} \nn \\
\eea

\no  Where we have used $i=-2(D-1)/3(D-2)$ and $j=D^2/(D-2)^2$. With the help of the general expressions obtained before, after inverting the operator $G$ we have the following propagator:
\bea
(G^{-1}) &=& \frac{2}{(\Box-m^2)}\left( P^{(3)}_{11}\right) - \frac{2}{m^2}\left( P^{(2)}_{11}\right) + \frac{2D(\Box-m^2)}{ (D+1)m^4}\left( P^{(1)}_{22}\right) \nn \\ \nn \\
&+& \frac{2\sqrt{D+1}}{(D+1)m^2}\left( P^{(1)}_{12} + P^{(1)}_{21} \right) \nn \\ \nn \\
&-&\frac{2}{3}\left[ \frac{(D-1)(D-2)\Box^2 - 2(D-1)(2D-1)\Box m^2 + 6D^2m^4 }{D^2(D+1)m^6}\right]  \left( P^{(0)}_{11} \right) \nn \\ \nn \\
&-&2 \left[ \frac{(D-1)^2(D-2)\Box^2 - 2(D-1)(D^2-D+1)\Box m^2 + D^2(D-2)m^4 }{D^2(D+1)m^6}\right] \left( P^{(0)}_{22} \right) \nn \\ \nn \\
&+& \frac{2}{3}\sqrt{3(D-1)} \left[ \frac{(D-1)(D-2)\Box^2 - (3D^2-4D+2)\Box m^2 + 3D^2 m^4 }{D^2(D+1)m^6}\right] \left( P^{(0)}_{12} + P^{(0)}_{21} \right) \label{SHP} \nn \\
\eea

\no Notice that there is a massive pole in the spin-3 sector and we have no dynamics in the lower spin sectors. In order to check that this spin-3 particle is, in fact, a physical particle let us analyze the sign of the imaginary part of the residue of the transition amplitude saturated in the sources. After taking the Fourier transformation,  in the momentum space, it is given by:
\bea
\mathcal{A_2}(k) = - \frac{i}{2} F_{\mu\nu\rho}^{*}(k)\left( G^{-1}(k)\right)^{\mu\nu\rho}_{\alpha\beta\gamma} F^{\alpha\beta\gamma}(k)
\eea

\no where $F_{\mu\nu\rho}(k)$ is the source in such space. We have a physical particle if the following condition on the transition amplitude is satisfied:
\bea
Im\left[ Res (\mathcal{A_2}(k))\mid_{k^2=-m^2} \right] > 0; \label{A}
\eea
otherwise, we would have a ghost at the spectrum. In our case we have: 
\bea
Im\left[ Res (\mathcal{A_2}(k))\mid_{k^2=-m^2} \right] = \lim\limits_{k^2 \xrightarrow{} -m^2} (k^2+m^2) \mathcal{A_2}(k) = \tilde{F}_{\mu\nu\rho}^{*}\tilde{F}^{\mu\nu\rho},
\eea
\no where $\tilde{F}_{\mu\nu\rho}=(P^{(3)}_{11})_{\mu\nu\rho}^{\alpha\beta\gamma}\tilde{F}_{\alpha\beta\gamma}$. As the spin-3 projection operator  ($P^{(3)}_{11}$)  is totally symmetric with  respect the indices $\mu\nu\rho$ and  $\alpha\beta\gamma$, traceless and transverse, the source term must have the same properties, i.e: 
\bea
\eta^{\mu\nu} \tilde{F}_{\mu\nu\rho} &=& 0 \\
k^\mu \tilde{F}_{\mu\nu\rho} &=& 0.
\eea
Once we have only one massive pole at the spin-3 sector we choose the frame given by  $k^\mu=(m,0,...,0_{(D-1)})$. Then we have:
\bea
k^\mu \tilde{F}_{\mu\nu\rho} = m\tilde{F}_{0\nu\rho} = 0 \Longrightarrow \tilde{F}_{0\nu\rho} = 0
\eea
\no which leave us only with the spatial contributions, given by:
\bea
Im\left[ Res (\mathcal{A_2})\mid_{k^2=-m^2} \right] = \tilde{F}_{ijl}^{*} \tilde{F}^{ijl}> 0 \quad;\quad i,j,k=1,...,D-1
\eea

From this spectral analysis, we can verify that the $SH$ model is free of ghosts, propagating only a spin-3 massive particle in $D$ dimensions. It is also interesting to notice that once the propagator of the $SH$ theory (\ref{SHP}) has a unique pole in the spin-3 sector, which goes with $1/(k^2+m^2)$ in the momentum space, the potential in the non relativistic limit (at least in $D=3+1$) between two sources exchanging spin-3 particles is a Yukawa potential $V(r)\sim e^{-m r}/r$, as also happens with the lower spin cases, namely Proca and $FP$ theories. 
The massless limit of higher spin theories however may have discontinuities, as observed by van Dam and Veltman for spin-2 theories \cite{VDV} and by Berends and Reisen \cite{Reisen} for the case of spin-3 theories.

\section{Hamiltonian positivity of the massless case}

The massless $SH$ theory is given by:
\be {\cal L}_{SH}^{m=0}= -\frac{1}{2}(\p_\alpha h_{\mu\nu\rho})^2+\frac{3}{2}(\p^{\mu}h_{\mu\nu\rho})^2+\frac{3}{2}(\p_{\alpha}h_{\nu})^2+ \frac{3}{4}(\p^{\nu}h_{\nu})^2-3\p_{\nu}h_{\rho}\p_{\mu}h^{\mu\nu\rho}.\ee

In order to study the Hamiltonian positivity of this system, we break manifest Lorentz covariance identifying terms with two time derivatives, which give us:
\bea {\cal L}_{SH}^{m=0}&=& \frac{1}{2}\,(\dot{h}_{ijk})^2 -\frac{3}{2}\,(\dot{\bar{h}}_{i})^2+\frac{1}{4}\,\dot{\lambda}^2+\frac{3}{2}\,\dot{\lambda}\,\p_i h_{00i}-3\,\p_k h_{0ij}\,\dot{h}_{ijk} +\frac{3}{2}\,\p_i  \lambda\, \dot{\bar{h}}_i\nn\\
&+& 3\,\p_i\bar{h}_0 \,\dot{\bar{h}}_i+ 6\,\p_i h_{0ij}\dot{\bar{h}}_j+ {\cal V}.\label{LV}\nn\\ \eea 

\no where we have used the following notation $\bar{h}_i=h_{ijj}$, $\bar{h}_0=h_{0kk}$ and $\lambda=h_{000}-3\bar{h}_0$, which is similar to the one adopted by \cite{Leonard}. Notice also that, all the terms without time derivatives are grouped on the symbol ${\cal V}$ which is explicitly given by:

\bea {\cal V}&=& -(\p_l \lambda)^2 - 3\,\p_l\lambda\p_l \bar{h}_0-\frac{3}{2}\,(\p_l \bar{h}_0)^2+\frac{3}{2}\,(\p_l h_{0ij})^2-\frac{1}{2}\,(\p_lh_{ijk})^2\nn\\
&+& \frac{9}{4}\,(\p_i h_{00i})^2- 3\,(\p_i h_{0ij})^2+ \frac{3}{2}\,(\p_i h_{ijk})^2+ \frac{3}{2}\,(\p_l \bar{h}_i)^2\nn\\
&+& \frac{3}{4}\,(\p_i\bar{h}_i)^2-3\,\p_lh_{00i}\p_l\bar{h}_i-\frac{3}{2}\,\p_ih_{00i}\p_l\bar{h}_l+ 3\,\p_j h_{k00}\p_i h_{ijk}\nn\\
&-&3\,\p_j \bar{h}_{k}\p_i h_{ijk}- 3\,\p_i h_{0ij}\p_j \lambda-6\,\p_j\bar{h}_0\p_i h_{0ij}
\eea

Calculating the conjugate momenta from (\ref{LV}) we have the following primary constraints $\varphi^i=\pi^{00i}\approx 0$ and $\varphi^{ij}=\pi^{0ij}\approx 0$. The consistency checking {\it a la} the Dirac-Bergman procedure of such constraints give us two additional secondary constraints $\chi^i$ and $\chi^{ij}$ which are given by:

\bea \chi^{i}&=& 3\left( \p^i \pi - \nabla^2\bar{h}^i+ \p_j\p_k h^{ijk}-\frac{1}{2}\p^i\p_j\bar{h}^j \right),\\
 \chi^{ij}&=& -3\left( \p_k \pi^{kij}+\frac{1}{2}\delta^{ij}\nabla^2 \lambda \right).\eea

\no One can verify that all the constraints are first class, which is related to the fact that the massless theory is gauge invariant under traceless reparametrizations i.e.: $\delta h_{\mu\nu\lambda}=\p_{(\mu}\tilde{\xi}_{\nu\lambda)}$ where $\tilde{\xi}_{\nu\lambda}=\tilde{\xi}_{\lambda\nu}$ and $\eta^{\nu\lambda}\tilde{\xi}_{\nu\lambda}=0$ . Then in the Lagrangian level, once we have two vectorial and two tensorial first class constraints, by considering the number of independent components of the totally symmetric rank-3 tensor one has $D(D+1)(D+2)/6-2D(D-1)/2-2(D-1)= (D-2)(D-3)(D+2)/6$ degrees of freedom which correspond  to the two helicities $+3$ and $-3$ in $D=4$. With these results in hand we then take the Legendre transformation, in order to obtain the canonical Hamiltonian, which can be written as:

\bea {\cal H}&=& \pi^2+\frac{1}{2}(\pi^{ijk})^2- \frac{3}{2D}(\pi^k)^2+\frac{3}{2D}\pi^k\p_k \lambda+ \left(\frac{5D-3}{8D}\right)(\p_k \lambda)^2\nn\\
&+&\frac{1}{2}\,(\p_l h_{ijk})^2-\frac{3}{2}\,(\p_i h_{ijk})^2+ 3\,\p_i h_{ijk}\,\p_j\bar{h}_k-\frac{3}{2}\,(\p_l\bar {h}_i)^2-\frac{3}{4}\,(\p_i\bar{h}_i)^2\nn\\
&+&h_{00i}\,\chi^i +h_{0ij}\,\chi^{ij},\eea

\no here the momentum $\pi$ is canonically conjugated to the combination $\lambda$. Following the same approach used by \cite{denisbend} one can now determine the partially reduced hamiltonian, which is done by using the secondary constraints as strong equalities, which give us:

\bea {\cal H}^{(pr)} &=& \pi_{ijk} \left[ \frac{1}{2}(P_{11}^{(3)}+P_{11}^{(2)})+\frac{(D-1)}{2D}P_{22}^{(1)}-\frac{\sqrt{D}}{2D}(P_{12}^{(1)}+P_{21}^{(1)})\right.\nn\\
&+&\left.\frac{(5D-7)}{6(D-1)^2}P_{11}^{(0)}+\frac{(D-2)(D-3)}{2(D-1)^2}P_{22}^{(0)}-\frac{(3D-5)}{6(D-1)^2}\sqrt{3(D-2)}(P_{12}^{(0)}+P_{21}^{(0)})\right]^{ijk}_{lmn}\pi^{lmn}\nn\\
&+&h_{ijk}\left[ \nabla^2 \left( -\frac{1}{2}P_{11}^{(3)}+\frac{(D-1)}{2}P_{11}^{(1)}\right)\right]^{ijk}_{lmn}h^{lmn},\eea

\no where we have written all the bilinear contractions of rank-3 tensors in terms of the spin-projection and the transition operators \footnote{Notice that we need to redefine our projectors, changing $D \to D-1$. Besides, the transverse and longitudinal operator are respectively given by: $\theta_{ij}=\delta_{ij}-\p_i\p_j/\nabla^2$ and $\omega_{ij}=\p_i\p_j/\nabla^2$.}. By using the secondary constraints one can then get rid of the terms proportional to $P_{11}^{(1)}$, $P_{11}^{(2)}$, $P_{22}^{(1)}$ and $(P_{12}^{(1)}+P_{21}^{(1)})$ separately. Notice also that, the spin-0 sector can be combined in order to vanish. This lead us with the reduced Hamiltonian given only in terms of the spin-3 projection operators, as expected for the massless spin $\pm 3$ particles described by the kinectic part of the $SH$ theory.

\be {\cal H}^{(r)}= \frac{1}{2}\pi_{ijk}\,\left( P_{11}^{(3)}\right) ^{ijk}_{lmn}\,\pi^{lmn}
-\frac{1}{2}h_{ijk}\,\left( \nabla^2P_{11}^{(3)}\right)^{ijk}_{lmn} h^{lmn}\geqslant 0\ee

\no once any projection operator sandwiched among two equal tensors have positive definite bilinear, and since $-\nabla^2$ has only positive eigenvalues, one can conclude that the reduced Hamiltonian is positive definite implying in its classical stability.
\section{Conclusion}

Here we provide a generalized set of spin-projection and transition operators for symmetric rank-3 tensors in $D$ dimensions. By setting $D=3+1$ we can recover the results of \cite{Reisen} and in this sense, we have a generalization of those projectors. In a work in progress, we are constructing a set of spin-projection operators to the case of non-symmetric fields in $D$ dimensions and in addition, we have considered the case of models with parity breaking in $D=2+1$, as it is the case of the self-dual models suggested by \cite{aragones31,ddhd}. 

The coefficients of the $SH$ model are determined with the help of the spin-projection operators, by projecting the equations of motion in the subspace of spin-3 and lower. We have noticed that we end up with two arbitrary coefficients $a$ and $t$. But both of them may be eliminated through field redefinitions. In a work in progress, we have been analyzing the particle content of the higher derivative models obtained at \cite{hds3}, but an additional difficulty is in the game now, once the models are gauge invariants we need to construct gauge-fixing terms, in order to obtain the propagator. Besides, as the models have higher derivatives, double poles should appear. It is interesting to say that in $D=2+1$ such higher derivative descriptions have lots of similarities with the new massive gravity model for spin-2 particles \cite{BHT}.  

Finally, we have given a detailed obtainment of the propagator of the $SH$ model by means of a general expression for bilinear lagrangians in terms of totally symmetric rank-3 tensors. From this analysis, we can verify that there is only a spin-3 particle in the spectrum of the theory. By adding a source term we have analyzed the sign of the imaginary part of the residue of the transition amplitude saturated and as it was expected the propagating mode is physical. In a way quite similar to the analysis performed by \cite{denisbend} we have also obtained the Hamiltonian  as well as the constraints of the massless $SH$ theory in $D$ dimensions. Thanks to the projection and transition operators we have then demonstrated that the system  is positive-definite, describing correctly the number of degrees as well as the spin.

\section{Acknowledgements}
This study was financed in part by the Coordenação de Aperfeiçoamento de Pessoal de Nivel Superior - Brasil (CAPES) - Finance Code 001. 
We acknowledge helpful discussions with Prof. Denis Dalmazi.

\section{Apendix}

Here we provide to the reader a detailed set of equations step by step where we write the terms in their bilinear form in terms of the spin-projection operators:
\begin{align}
h_{\mu\nu\rho} \Box h^{\mu\nu\rho} = & h_{\mu\nu\rho}\left[ \Box \left( P^{(3)}_{11} + P^{(2)}_{11} + P^{(1)}_{11} + P^{(1)}_{22} + P^{(0)}_{11} + P^{(0)}_{22} \right)^{\mu\nu\rho}_{\kappa\lambda\sigma} \right] h^{\kappa\lambda\sigma} \nn \\
h_{\mu\nu\rho} \p^\mu \p_\alpha h^{\alpha\nu\rho} = & h_{\mu\nu\rho}\left[ \frac{\Box}{18} \left( 6 P^{(2)}_{11} + 12 P^{(1)}_{22} + 6 P^{(0)}_{11} + 18 P^{(0)}_{22} \right)^{\mu\nu\rho}_{\kappa\lambda\sigma} \right] h^{\kappa\lambda\sigma} \nn \\
h_{\nu} \p^\nu \p_\mu h^{\mu} = & h_{\mu\nu\rho}\left[ \frac{\Box}{9} \left( 3(D-1) P^{(0)}_{11} + 9 P^{(0)}_{22} + 3\sqrt{3(D-1)}( P^{(0)}_{12} + P^{(0)}_{21}) \right)^{\mu\nu\rho}_{\kappa\lambda\sigma} \right] h^{\kappa\lambda\sigma} \nn \\
h_{\nu} \Box h^{\nu} = & h_{\mu\nu\rho}\left[ \frac{\Box}{9} \left( 3(D+1) P^{(1)}_{11} + 3 P^{(1)}_{22} + 3(D-1) P^{(0)}_{11} + 9 P^{(0)}_{22} \right)^{\mu\nu\rho}_{\kappa\lambda\sigma} \right] h^{\kappa\lambda\sigma} \nn \\
+ & h_{\mu\nu\rho}\left[ \frac{\Box}{9} \left( 3\sqrt{D+1}( P^{(1)}_{12} + P^{(1)}_{21}) + 3\sqrt{3(D-1)}( P^{(0)}_{12} + P^{(0)}_{21}) \right)^{\mu\nu\rho}_{\kappa\lambda\sigma} \right] h^{\kappa\lambda\sigma} \nn \\
h_{\rho} \p_\mu \p_\nu h^{\mu\nu\rho} = & h_{\mu\nu\rho}\left[ \frac{\Box}{18} \left( 6 P^{(1)}_{22} + 18 P^{(0)}_{22} + 3\sqrt{D+1}( P^{(1)}_{12} + P^{(1)}_{21}) \right)^{\mu\nu\rho}_{\kappa\lambda\sigma} \right] h^{\kappa\lambda\sigma} \nn \\ 
+ & h_{\mu\nu\rho}\left[ \frac{\Box}{18} \left( 3\sqrt{3(D-1)} ( P^{(0)}_{12} + P^{(0)}_{21}) \right)^{\mu\nu\rho}_{\kappa\lambda\sigma} \right] h^{\kappa\lambda\sigma} \nn \\
m^2 h_{\mu\nu\rho} h^{\mu\nu\rho} = & h_{\mu\nu\rho}\left[ m^2 \left( P^{(3)}_{11} + P^{(2)}_{11} + P^{(1)}_{11} + P^{(1)}_{22} + P^{(0)}_{11} + P^{(0)}_{22} \right)^{\mu\nu\rho}_{\kappa\lambda\sigma} \right] h^{\kappa\lambda\sigma} \nn \\
m^2 h_{\mu} h^{\mu} = & h_{\mu\nu\rho}\left[ \frac{m^2}{9} \left( 3(D+1) P^{(1)}_{11} + 3 P^{(1)}_{22} + 3(D-1) P^{(0)}_{11} + 9 P^{(0)}_{22} \right)^{\mu\nu\rho}_{\kappa\lambda\sigma} \right] h^{\kappa\lambda\sigma} \nn \\
+ & h_{\mu\nu\rho}\left[ \frac{m^2}{9} \left( 3\sqrt{D+1}( P^{(1)}_{12} + P^{(1)}_{21}) + 3\sqrt{3(D-1)}( P^{(0)}_{12} + P^{(0)}_{21}) \right)^{\mu\nu\rho}_{\kappa\lambda\sigma} \right] h^{\kappa\lambda\sigma} \nn
\end{align}
\no Collecting all these terms one can write the first and most tedious term of the lagrangian density given by (\ref{lagop}).

\end{document}